\documentstyle[12pt,aasms4]{article}

\slugcomment{Accepted for publication in the Astrophysical Journal}

\lefthead{}
\righthead{}

\begin{document}
\title
{\bf Interaction of Rayleigh-Taylor Fingers and Circumstellar Cloudlets
in Young Supernova Remnants}{}{}

\author{Byung-Il Jun and T.W. Jones}
\affil{Department of Astronomy, University of Minnesota \\
116 Church Street, S.E., Minneapolis, MN 55455}
\and 

\author{Michael L. Norman} 
\affil{Laboratory for Computational Astrophysics \\ National
Center for Supercomputing Applications \\ 
Department of Astronomy, University of Illinois at Urbana-Champaign}

\begin{abstract}

   We discover a new dynamical mechanism that significantly enhances 
the growth of Rayleigh-Taylor
fingers developed near the contact interface between the supernova
ejecta and swept-up ambient gas in young supernova remnants if  
the supernova remnant expands into a clumpy (cloudy) circumstellar medium.
Our numerical simulation demonstrates that
large Rayleigh-Taylor fingers
can obtain a sufficient terminal velocity to protrude through the forward
shock front by taking extra kinetic energy from vorticies generated by
shock-cloud interactions.  We suggest this mechanism as a means to
generate the aspherical expansion of the supernova ejecta.
Ambient magnetic fields
are stretched and amplified as the Rayleigh-Taylor fingers protrude,
possibly leading to strongly enhanced radio emission.  The material in
the protrusions originates from the ejected stellar material
with greatly enhanced heavy elements. Therefore, it can be a strong
X-ray emitter.  The timescale for the Rayleigh-Taylor fingers to
reach the forward shock depends on
the size, mass density and distribution of clouds being engulfed by
the supernova shock, although the details will require further
numerical investigation.

\end{abstract}

\section{Introduction} % 1

A very-long-baseline interferometric (VLBI) radio image of supernova
1986J shows a shell of emission with protrusions (\cite{bar91}).
These protrusions extend from the center to twice the radius of the
shell.  The shell boundary is commonly associated with the forward
shock of the ejecta.  The Cas A also shows several bow-shock features
that are thought to be the results of high-velocity clumps
(\cite{bgp87}). 
These bow-shock features serve as clear examples of aspherical expansion of
supernova ejecta.  Deviation from spherical symmetry in the expansion
of supernova 
remnants could be caused by a number of mechanisms.  Asymmetric
explosion of the supernova itself coming from an early instability such as
Rayleigh-Taylor (R-T) instability of the stellar core/envelope
(\cite{fma91}) might result in these 
protrusions, as well as the bow-shock features in Cas A.
Several V-shaped
features in the X-ray image of the Vela supernova remnant have been
proposed to be the result of fragments ejected during the formation of
the neutron star (\cite{lrs95}).  
Loeb et al. (1995) suggested that
fragmentation may occur during gravitational 
collapse as a result of both convective and rotational instabilities.
On the other hand, Gull (1975) suggested that the fast moving
knots in Cas A could be produced by the R-T instability 
in the decelerating supernova remnant shell (different
from the early R-T instability inside the exploding star).
Numerical simulations show that the growth of R-T fingers in supernova
remnant shells expanding into a uniform ambient medium cannot reach the
forward shock front (\cite{cbe92,jn96a,jn96b}). 
The investigation of the shell R-T instability has been extended to include the
effect of radiative cooling (\cite{che95}), because
radiative cooling is expected to increase the density contrast in the
unstable region, and, therefore, the growth of the R-T fingers.  However,
it turns out that even then the unstable fingers are not able to form the
protrusions through the forward shock.   
Recently, Blondin, Lundqvist, and Chevalier (1996) carried out numerical
simulations of the interaction of a supernova with an axisymmetric
density distribution which has a high density in equator.
They found that a protrusion appears along the
axis if the angular density gradient exceeds a moderate value.
This model is able to generate two oppositely directed polar
protrusions.
However, another mechanism is still required to produce
more than two protrusions.

 We propose that multiple protrusions from a shell in supernova remnants
can be produced through the interaction of the shell with a
clumpy (cloudy) medium.  The existence of such clumps have been
invoked by many authors to explain various observations.  For example,
the interaction of ejecta with a clumpy wind model has
been proposed as the origin of the broad and intermediate-width lines
in the spectrum of the peculiar supernova SN 1988Z (\cite{chu94}).
Chugai and Danziger also identify the optical
emission of SN 1986J with the radiation from the shocked dense
component of the clumpy presupernova wind.  They argue that the clumpy
circumstellar medium model is more attractive than the clumpy ejecta
model in SN 1986J because the early optical spectra do not provide any
evidence for clumpiness in the outer layers of the envelope.
The evolution of supernova remnants in a clumpy
ambient medium has been studied by one-dimensional numerical
simulation by several authors (\cite{shi78,dic89,dic93}).
A two-dimensional study of this model has
recently been carried out to explore the strong radio emission from the shell
(\cite{jun95}).  In this {\it Letter}, we illustrate that shell protrusions
can be produced by the enhanced growth of R-T fingers as a result of
their interactions with a clumpy medium.  Our model does not require an
aspherical explosion of the supernova.
The key physics of our model is the fluid dynamics of engulfed clouds in the
intershock region.  The primary effect of a cloudy medium on the
R-T fingers is to disturb and disrupt them.  However, we discovered
an additional effect, which is that vorticity generated by
shock-cloud interactions can enhance the growth of large R-T fingers
through a collective effect, which we illustrate through numerical
simulations.

\section{Limited Growth of R-T Fingers in a Uniform Medium Model} %2

In this section, we discuss the necessary condition for the R-T fingers to
penetrate the outer shock and ask why the R-T fingers in a uniform
medium model could not produce the protrusions.
Generally, if the terminal velocity of the R-T finger is greater than
$ v_e = {1 \over 4}v_s$ (the postshock fluid speed for a strong
$\gamma=5/3$ adiabatic shock) where $v_e$ is the escape velocity and
$v_s$ is the shock velocity, the R-T finger can protrude through
the shock front.  The terminal velocity of the finger $v_t = \sqrt{ {4
\over 3} {\rho_f \over \rho_b} gr}$ can be
determined roughly by equating the ram pressure with the effective
gravitational force (\cite{cow75}). Here $\rho_f$ is the finger density,
$\rho_b$ is the background density, $g$ is the effective gravity, and
$r$ is the finger radius.  Therefore, a larger and denser finger
reaches a higher terminal velocity for a given effective gravity.
The terminal velocity of the R-T finger can be written as a function
of mass density alone for a given effective gravity using mass
conservation. The result is $v_t \propto \rho_f^{1/3} g^{1/2}$.  Therefore,
as studied by Chevalier and Blondin (1995), contraction of the
finger by radiative cooling will increase the terminal velocity.

   Numerical simulations (\cite{cbe92,jn96a,jn96b}) show
that even the largest R-T finger formed in a adiabatic shell expanding into
a uniform medium cannot reach a
sufficient terminal velocity to penerate the shock front.    
Figure 1a (no cooling included) shows that the largest R-T fingers
extend through only about half of the intershock region in a uniform
medium model. 
This limited extension of the R-T fingers is due to
the drag and the divergence of the volume in spherical geometry 
(\cite{cbe92}). 
The finger expands faster than its environment as it moves outward and
the expanded finger will reach a lower terminal
velocity since, under these cirsumstances, $v_t \propto r^{-1}g^{1/2}$.
Also, the finger sheds its mass due to the
Kelvin-Helmholtz instability resulting in a lower terminal velocity.
Even in the presence of radiative cooling, the R-T
fingers did not produce protrusions penetrating the outer shock (\cite{che95}).
These numerical results suggest that in order for the finger
to reach a sufficient velocity to achieve penetration, additional
kinetic energy needs to be 
added to the finger by some new mechanism.  We have discovered that
the vorticity created by multiple shock-cloud interactions provides
such a mechanism.

\section{Numerical Simulation of SNR in a Clumpy Medium} %3

  As a number of authors have reported, the interaction between a shock
wave and a dense cloud results in cloud fragmentation and
turbulent mixing through a combination of the Kelvin-Helmholtz and the
Rayleigh-Taylor instabilities (\cite{sto92,jon93,kle94,mac94}).
These simulations assume a single isolated cloud and a uniform
post-shock state.  However, the interaction of a R-T unstable
supernova remnant shell with multiple clouds is considerably more
complex.  For example, refraction of the shock wave by the clouds
leads to enhanced 
transverse mixing through a collective effect (\cite{nor88}).
We study the collective effect of the shock-cloud interactions on
the expansion of supernova remnants and the growth of the R-T fingers
by using a multi-dimensional MHD code (ZEUS-3D).  

\subsection{Initial Condition}

   In our simulation the initial density profile of
the ejected supernova material follows a power law with $\rho \sim
r^{-7}$ in the outer 3/7 of the star's mass and a constant density in
the inner 4/7 of the star's mass.  An initial energy of $10^{51} ergs$
is deposited as kinetic energy.  The initial condition of the
supernova is the same as the one in Jun and Norman (1996a).   A number of small
clouds (76) are distributed randomly in the ambient medium of density $1.67
\times 10^{-24} g/{c.c}$.  The constant
density contrast between the cloud and intercloud medium is chosen to
be 5.  The clouds are given a circular cross-section shape 
with constant radius, $ r_c = 0.05 pc$.  
A $3.5 \mu G$ ambient magnetic field is assumed to be tangential to the shock
front and lying in the computational plane.
The computation is carried out in a two-dimensional wedge of $\pi /4 $
radians in
spherical polar geometry.  The computational plane is resolved
uniformly with 1200 zones in the radial direction and 600 zones in the
angular direction.
After 500 years ($\mu = 4$ where $\mu$ is the mass ratio defined by the
swept-up mass divided by the ejecta mass) , about 51 clouds are
encountered by the shock front.  The filling fraction of the area
occupied by the clouds at this time is only about $8.6 \%$.  Our
choice of initial conditions was 
originally aimed at the study of young supernova remnants
a few hundred years of age. 
Therefore, much smaller clouds in a circumstellar medium with a power
law background density, $\rho \propto r^{-2}$, would be more appropriate to
model the early protrusions in the remnants of Type II supernova.
However, the qualititive description of the 
mechanism will remain the same.

\subsection{Results}

   The dynamical evolution of the intershock region in a clumpy medium
model can be characterized by three stages (\cite{jun95}).
They are the development of the R-T instability, the interaction between the
shock and the overrun clouds, and the collective effect of the clouds engulfed
by the shock in the intershock region.
First, the R-T instability driven by the deceleration of the ejecta
develops near the contact discontinuity with the shocked ambient gas.
The nonlinear growth of the
R-T instability has been studied in detail in 2D and 3D (\cite{jn96a,jn96b}).
As the first cloud encounters the supernova shock,
it is compressed by the shock transmitted into the cloud.  The
incident shock is also reflected into the intercloud region forming a
bow-shock (\cite{mck75}).  This bow-shock
is much weaker than the original shock giving in our simulation 
a compression ratio about 2.  
The Kelvin-Helmholtz instability develops
at the cloud surface due to the sheared flow
between the cloud and the bow shock.  When the cloud
shock reaches the rear of the cloud, a rarefaction wave is reflected
back into the cloud leading to a reexpansion of the shocked cloud
(\cite{mck88}). Now the cloud is destroyed by combined instabilities.
When the shocked clouds are subsequently overrun by the contact discontinuity, 
the R-T fingers formed from the SNR contact discontinuity are disrupted by
the ensuing collisions and the SNR reverse shock is severely deformed.  The
disrupted clouds 
interact strongly with the supernova ejecta, and the flow in the intershock
region becomes turbulent.  The complexity of the intershock flow
increases as the formed shock overruns more clouds.   

  As the
supernova remnant expands, the intershock region broadens accordingly
and the direct impacts of cloud collisions on the global structure of
the shell are less important. That is
because the
shocked clouds are disrupted more completely due to the longer travel time
before they reach the contact discontinuity from the forward shock.
Then large R-T fingers can develop without
much disturbance from the clouds,a because the wavelength is much larger
than the cloud.   Locally, however, shock-cloud interactions maintain
a higher level of shell turbulence than in the uniform medium model
(compare Figs. 1a and 1b).
The developed R-T fingers now interact with a number of
eddies generated from the shock-cloud collisions.  The motion of R-T
fingers can be deflected as they encounter strong eddies.  As a
result, the growth of R-T fingers can either be aided or
hindered, depending on the sense of eddy rotation.

Figure 1b shows
the density image of a supernova remnant in a clumpy medium when the
remnant swept up 4 times its own mass (t=500 years).  The
first obvious result is that the entire intershock region is turbulent.
Both the forward shock and the reverse shock are
deformed by collisions with the clouds. 
The ``protrusion'' from the shell is just forming out of a large
R-T finger (see also Figure 2 which shows the protruding finger at
t=650 years). 
We observed the formation of strong vortices 
on the right side of this finger due to successive shock-cloud
interactions and eddy mergers (We discuss this mechanism in more detail in
section 4.).   Note that only one finger out of several in this
simulation will lead to a ``protrusion''.  So, the mechanism is
expected to generate only a few protrusions over the entire blast wave.
Figure 1c shows that strong magnetic fields
around this long R-T finger extended up to the forward shock front.
This strong magnetic field structure will likely stand out through enhanced
radio emission if an energetic electron population is accelerated by
the shocks or incorporated from the external medium.
To establish the composition of material in the protrusion, we can examine
the distribution of the mass fraction, which is a passively advected
quantity (\cite{jns95,jn96a}) and allows us to
identify ejected stellar material.  
Figure 1d represents the
distribution of the mass fraction in the intershock region where the black
color corresponds to the ejected stellar material.  It is clearly illustrated
that the material in the core of protrusion is composed of the supernova
ejecta.   Since that should include heavy elements, this region
should be a strong X-ray emitter.

\section{Discussion of Protrusion Mechanism} %4

Strong vorticies are generated by the velocity shear
between the cloud shock and the intercloud shock.  Then the
flow speed at the edge of the vortex $v_w$ is approximately $v_s - v_c$
where $v_c$ is the cloud shock velocity and is given as $({\rho_b \over
\rho_c})^{1/2} v_s$ where $\rho_c$ is the mass density of cloud.   
This linear velocity can be as high as
$v_s( 1- ({\rho_b \over \rho_c})^{1/2})$.  It can be
higher than the escape velocity, $v_e$, if ${\rho_c \over \rho_b} >
{16\over 9}$.   Therefore, a disrupted cloud will generate a very
strong vortex.  The interaction of the R-T finger with a single, strong
vortex can enhance the growth of the R-T finger significantly.
This enhancement can be even greater if the interaction occurs with a
larger vortex generated by vortex-vortex mergers.
The interaction between two eddies in two dimensional flow
depends on the sign of vorticity.  
Eddies of like vorticity (same sign of vorticity vector)
will tend to merge with each other while eddies of opposite sign of
vorticity will tend to avoid each other.  
Vortex interactions in three dimensions is considerably more complex
than this, with the result that eddy mergers may occur less
frequently in three dimensions.  However, eddy mergers are not
essential to produce the protrusions. The critical point for the R-T
finger to achieve a sufficient terminal velocity is the efficient
transfer of rotational energy to the R-T finger.
Whenever two adjacent eddies have opposite signs of vorticity, they can allow a
flow channel to form between them.  Therefore, once the R-T fingers enter a
channel with the flow directed towards the forward shock, 
the moving finger obtains more kinetic energy and moves faster.
The R-T finger can be further boosted by a successive interaction with
vorticies and therefore produce the ``protrusion'' as depicted in Figure 3.   
In our model, additional kinetic energy for
the protrusion originates from the outer shock.  The cloud plays a
role as a mediator that transfers the kinetic energy of
the outer shock (blast wave) into the R-T fingers. The extracted
kinetic energy from the blast wave by the cloud encounters results in
the retarded propagation of the blast wave, therefore satisfying energy
conservation.

  In our idealized numerical simulation, the size and density of clouds
are assumed as constants throughout the space.   In order to identify
the roles played by
the size and mass density of clouds in boosting the
growth of R-T finger, we should consider kinetic energy conservation
between the finger and vortex (rotating material formed by a disrupted
cloud).   
The total kinetic energy of the finger
and cloud can be written as a sum of kinetic energy of the finger and
the rotational energy of the vortex
\begin{equation}
E_t = {1\over 2} I w^2 + {1 \over 2} M_f v_t^2
\end{equation}
where $I$ is the moment of inertia of the vortex, $w$ is the angular
velocity of the vortex, $M_f$ is the finger mass, and $v_t$ is the
terminal velocity of the finger.   The rotational energy of the cloud
vortex can 
also be approximated as ${1\over 2} \delta M_c v_w^2$ where $\delta M_c$ is
the mass of cloud vortex that rotates with the velocity $v_w$.  
Now one can derive the final velocity of the finger ($v_f$)
after the cloud vortex has merged into the finger and transfered its
rotational kinetic energy,
\begin{equation}
v_f^2 \approx {1 \over M_f + \delta M_c} (M_f v_t^2 + \epsilon \delta
M_c v_w^2)
\end{equation}
where $\epsilon$ is the energy transfer efficiency between the finger
and the cloud vortex.
To produce the protrusion, the final velocity has to exceed the escape
velocity $v_e$.    Assuming that the finger is at rest before it
interacts with the cloud vortex (just to consider the effect of the
cloud vortex), one can get the condition for the protrusion as
functions of the cloud mass and density
\begin{equation}
\epsilon {\delta M_c \over M_f + \delta M_c}( 1 - ({\rho_b \over
\rho_c})^{1/2}) ^2 > {1 \over 16}.
\end{equation}
Therefore, one can see that the final velocity of the finger increases
as the mass and density of the cloud increases.
However, it should be
noted that the size and mass density of the cloud are critical quantities in
determining the cloud disruption timescale that will be required to
form the vortex.   For larger and denser clouds, the disruption
takes longer.
According to Klein
et al. (1994), small clouds are destroyed in several cloud crushing
times, where the cloud crushing time $t_{cc}$ is defined as the time for the
shock to cross through the cloud.  
By taking the parameters used in our simulation, we
obtain a cloud crushing time, $t_{cc} \approx 20 years$.   Therefore,
the clouds are expected to be destroyed within a few decades.
That is in a good agreement with the results in our
simulation, where individual clouds took about 100 years to be completely
destroyed.   Allowing time for the R-T finger to travel up
to the outer shock front, it is expected to take a few hundred years
before the R-T finger protrudes through the shock front.  This is why the
protrusion of the finger appears in the later stage of the evolution
in our simulation.    Although denser and larger clouds can play a
bigger role in transfering the rotational energy to the finger, the
cloud needs be disrupted before it interacts with the R-T finger,
which sets the upper limit of the density and size of a cloud that can
participate in this process.
In addition, the distribution of clouds is also important, since the
shocked cloud can affect the growth of the R-T finger negatively as
well as positively.  In order to understand the detailed dynamical
picture, further numerical investigation will be required.

The ultimate fate of the
protrusion is unknown as yet, because we could not evolve the
remnant much longer due to the limited
computational domain.  Cowie (1975) 
predicted that the penetrating R-T finger can
be re-overtaken by the shock wave due to the increased ram pressure
encountered by the finger once it breaks through the shock.
He conjectured that the fragments will oscillate in and out
of the blast wave.    However, this oscillation process is questionable.
We speculate that a secondary shock may be
produced by the increased ram pressure and transmitted into the R-T
finger as the finger protrudes through the outer shock (blast wave).  
The finger will begin to be destroyed by Rayleigh-Taylor and
Kelvin-Helmholtz instabilities.  This process will be very similar 
to the dynamical evolution of a cosmic bullet studied by Jones, Kang,
and Tregillis (1994).   As the finger is destroyed by the combined
instabilities, it will be decelerated and be overtaken by
the outer shock.  Now, the bow-shock produced by the protruding
finger may become flattened by the stabilizing effect of the strong
shock.   This recovery process may take a long time if the density
and velocity of the protruding R-T finger is large. 
Also, the evolution of the remnant in a wind-blown ambient medium,
where the density is proportional to $r^{-2}$, may result in a
different protrusion dynamics; possibly a stronger protrusion as seen
in SN 1986J.
The dynamics of the protrusions will be of great
importance in explaining observed structures, 
since it may be strongly associated with emission processes.
As the material in the protruding finger disperses, the emission
(radio, X-ray, and possibly optical) will fade away. 
It is likely that this process will control the timescale of the
disappearance of the bow-shock features in Cas A identified by Braun et
al. (1987). For example, the opening angle of the bow-shock features
will become narrower or wider depending on the evolutionary stage.

\section{Conclusions}

  We have found that the interaction of R-T fingers and
circumstellar cloudlets in young supernova remnants can enhance the
growth R-T fingers significantly and generate protrusions from the
shell, as the large R-T finger distorts the outer shock front.
Some R-T fingers obtain extra kinetic energy
through their successive interactions with vortices generated by the
shocked clouds.  Therefore, extra kinetic energy for the protrusions
originates from the kinetic energy of the blast wave, and the cloud plays a
role as a mediator between the outer shock and the R-T fingers.
The protrusion process takes place after the
intershock region becomes turbulent due to cloud disruptions.  The
timescale for the protrusions is sensitive to the size and mass density of
the cloud as well as the number density  and distribution of clouds in
the circumstellar medium.   Our model suggests that the origin of
bow-shock features in Cas A could be protruding R-T fingers.
We also suggest that our mechanism can be a possible origin for
protrusions in remnants such as SN 1986J while much smaller clouds in
the circumstellar medium will be required to produce protrusions in
the immediate circumstellar environment.
Our model suggests that the protrusions may be the
sites of strong radio and X-ray emission.
Further simulations are needed to study the effects of
three-dimensionality, gradients in the ambient medium, and different
cloud properties on the dynamics of shell protrusions.

\acknowledgements

  T.W.J and B.-I.J. were partly supported at the University of
Minnesota by NSF grant AST-9318959 and by the Minnesota
Supercomputer Institute.

\clearpage

\figcaption[fig1.ps]{Grey scale images of two simulated supernova remnants at
$\mu = 4$ (500 
years) where $\mu$ is the ratio of the swept-up mass and
the ejecta mass. (a) gas
density of a supernova remnant evolving in a uniform ambient medium.
(b) gas density of a supernova remnant evolving in a cloudy medium. (c) the
magnetic field strength in the cloudy medium model. (d) mass fraction
distribution in the cloudy medium (black is stellar ejecta). \label{fig1}}

\figcaption[fig2.ps]{Grey scale images of density (top panel) and
magnetic field strength (bottom panel) of a simulated supernova remnant in a
cloudy medium at t=650 years.  \label{fig2}}

\figcaption[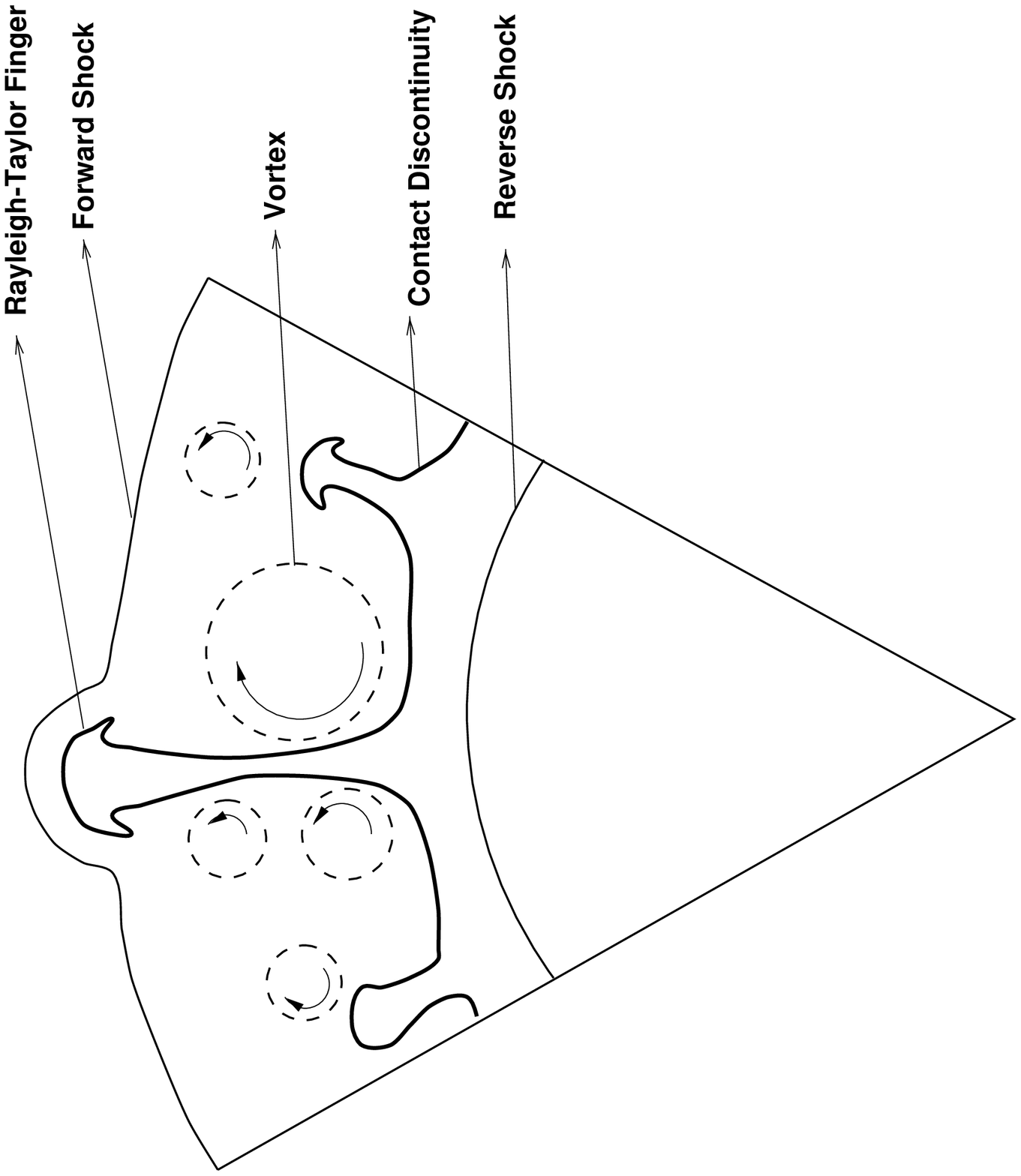]{A schematic representation of the proposed
mechanism for the 
formation of protrusions in a supernova shell.   Large vortices are 
created in the shell by shock-cloud interactions.  A dense
Rayleigh-Taylor finger can be ``channeled'' toward the forward shock
through counter-rotating vortices of the correct sign. \label{fig3}}

\end{document}